\documentclass[jgr,draft]{agu2001}

\usepackage{amsfonts}
\usepackage{amsmath}
\usepackage{graphicx}

\newcommand{\coloneqq}{\ensuremath{\mathrel{:=}}}

\newcommand{\Kb}{\emph{K. brevis }}
\newcommand{\bfx}{\mathbf{x}}
\newcommand{\bfu}{\mathbf{u}}

\authorrunninghead{OLASCOAGA ET AL.}
\titlerunninghead{Tracing the early development of harmful algal blooms with the aid
of Lagrangian coherent structures}
\lefthead{OLASCOAGA ET AL.}
\righthead{TRACING THE EARLY DEVELOPMENT OF HAB\lowercase{s}}
\journalid{2007} \articleid{X}{X}
\paperid{2007GLxxx} \cpright{AGU}{2007} \received{\today}
\revised{\today} \accepted{\today} 

\authoraddr{M.~J. Olascoaga, F.~J. Beron-Vera and L.~E. Brand,
RSMAS/AMP, University of Miami, 4600 Rickenbacker Cswy.,
Miami, FL 33149, USA. (jolascoaga@rsmas.miami.edu)}
\authoraddr{H. Ko\c{c}ak, Departments of Computer Science and
Mathematics, University of Miami, 1365 Memorial Dr., Coral
Gables, FL 33124, USA.}

\begin{document}

\title{Tracing the Early Development of Harmful Algal Blooms on
the West Florida Shelf with the Aid of Lagrangian Coherent
Structures}

\authors{
M.~J. Olascoaga,\altaffilmark{1} F.~J.
Beron-Vera,\altaffilmark{1} L.~E. Brand,\altaffilmark{1} and H.
Ko\c{c}ak\altaffilmark{2}}

\altaffiltext{1}{Rosenstiel School of Marine and Atmospheric
Science, University of Miami, Miami, Florida, USA.}
\altaffiltext{2}{Departments of Computer Science and
Mathematics, University of Miami, Miami, Florida, USA.}

\begin{abstract}
Several theories have been proposed to explain the development
of harmful algal blooms (HABs) produced by the toxic
dinoflagellate \emph{Karenia brevis} on the West Florida Shelf.
However, because the early stages of HAB development are
usually not detected, these theories have been so far very
difficult to verify. In this paper we employ simulated
\emph{Lagrangian coherent structures} (LCSs) to trace the early
location of a HAB in late 2004 before it was transported to an
area where it could be detected by satellite imagery, and then
we make use of a population dynamics model to infer the factors
that may have led to its development. The LCSs, which are
computed based on a surface flow description provided by an
ocean circulation model, delineate past and future histories of
boundaries of passively advected fluid domains. The population
dynamics model determines nitrogen in two components, nutrients
and phytoplankton, which are assumed to be passively advected
by the simulated surface currents. Two nearshore nutrient
sources are identified for the HAB whose evolution is found to
be strongly tied to the simulated LCSs. While one nutrient
source can be associated with a coastal upwelling event, the
other is seen to be produced by river runoff, which provides
support to a theory of HAB development that considers nutrient
loading into coastal waters produced by human activities as a
critical element. Our results show that the use of simulated
LCSs and a population dynamics model can greatly enhance our
understanding of the early stages of the development of HABs.
\end{abstract}

\begin{article}

\section{Introduction}

The largest and most frequent harmful algal blooms (HABs)
caused by the toxic dinoflagellate \emph{Karenia brevis} tend
to occur along the southern portion of the West Florida Shelf
(WFS) \citep{Steidinger-Haddad-81,Kusek-etal-99,
Brand-Compton-07}. Associated with typical HAB events is
widespread death of sea life resulting from the brevetoxins
produced by \Kb \citep{Bossart-etal-98,
Landsberg-Steidinger-98,Landsberg-02,Shumway-etal-03,
Flewelling-etal-05}.Brevetoxins are also known to cause injury
to humans through ingestion of contaminated seafood, skin
contact, or inhalation of aerosolized brevetoxins in coastal
regions \citep{Backer-etal-03,Kirkpatrick-etal-04} .

Dinoflagellates have slow growth rates
\citep{Brand-Guillard-81}. As a result, a necessary condition
for achieving a high dinoflagellate population density is that
physical processes (stretching and folding of fluid elements)
do not act to significantly reduce the population density
during the growth phase. Indeed, dinoflagellate blooms often
occur in enclosed basins. The tendency of \Kb to cause large
HABs along the southern WFS may then be explained by the
presence of a cross-shelf transport barrier that has been
revealed in the analysis of drifter trajectories
\citep{Yang-etal-99} and in an appropriate synthesis of modeled
surface currents \citep{Olascoaga-etal-06} . In
\citet{Olascoaga-etal-06} it was hypothesized that the presence
of such a transport barrier, which was characterized as a
\emph{Lagrangian coherent structure} (LCS), provides favorable
conditions for the development of HABs in the southern WFS by
inhibiting cross-shelf transport and thereby allowing for a
greater nutrient buildup. We remark that, being of a
fundamentally Lagrangian nature, the aforementioned transport
barrier could not be identified visually from snapshots of the
simulated surface velocity field.

Satellite ocean color sensors have demonstrated the ability to
provide a very useful HAB monitoring capability
\citep{Tester-etal-98,Stumpf-etal-03,Tomlinson-etal-04,Hu-etal-05,
deAraujo-etal-07}. To be detectable by current satellite
sensors, however, the \Kb cell concentration must be of the
order of 10$^5$ cell l$^{-1}$ (equivalently 1 mg Chl m$^{-3}$)
\citep{Tester-etal-98}, which may take a period of about one
month from background concentration to be attained
\citep{Tester-Steidinger-97,Steidinger-etal-98}. As a
consequence, satellite ocean color imagery by itself is not
capable of detecting HAB initiation, and hence cannot be used
to determine what factors favor HAB development, which is a
subject of continuous debate.

Earlier work \citep{Rounsefell-Nelson-66} has associated the
development of HABs on the WFS with river runoff, pollution,
and other nutrient enriching phenomena produced by human
activities. Consistent with this earlier work and the increase
in the frequency of HABs experienced during the last century, a
recent study \citep{Brand-Compton-07} has indicated that \Kb
concentrations are indeed higher near the shoreline than
offshore, and that nutrient-rich freshwater runoff from land is
likely to be a major source of nutrients for the development of
HABs. Other investigators \citep{Hu-etal-06} have argued that,
in addition to river runoff, submarine groundwater discharge,
particularly when enhanced by an active hurricane season,
should play a role in stimulating HABs on the WFS. It has been
also suggested \citep{Tester-Steidinger-97} that some HABs
initiate several tens of kilometers offshore, near frontal
regions, and that upwelling events provide a mechanism for
supplying the required nutrients for their stimulation. One
further, more complicated, theory has been proposed
\citep{Lenes-etal-01,Walsh-Steidinger-01,Walsh-etal-03,Walsh-etal-06}
which explains HAB development as a sequence of events preceded
by iron-rich Saharan dust deposition. In short, the theory
assumes that the Saharan dust stimulates \emph{Trichodesmium}
blooms, which fix nitrogen. The blooms decompose on sinking and
release dissolved organic nitrogen, which stimulates a \Kb seed
population near the bottom. Coastal upwelling is then required
to help this population reaching the surface, where light
inhibition is alleviated and a HAB finally develops. However,
as already pointed out, the early stages of the development of
HABs are usually not detected, which makes it very difficult to
test the aforementioned theories.

\begin{figure}[t]
\centering
\includegraphics[width=27pc]{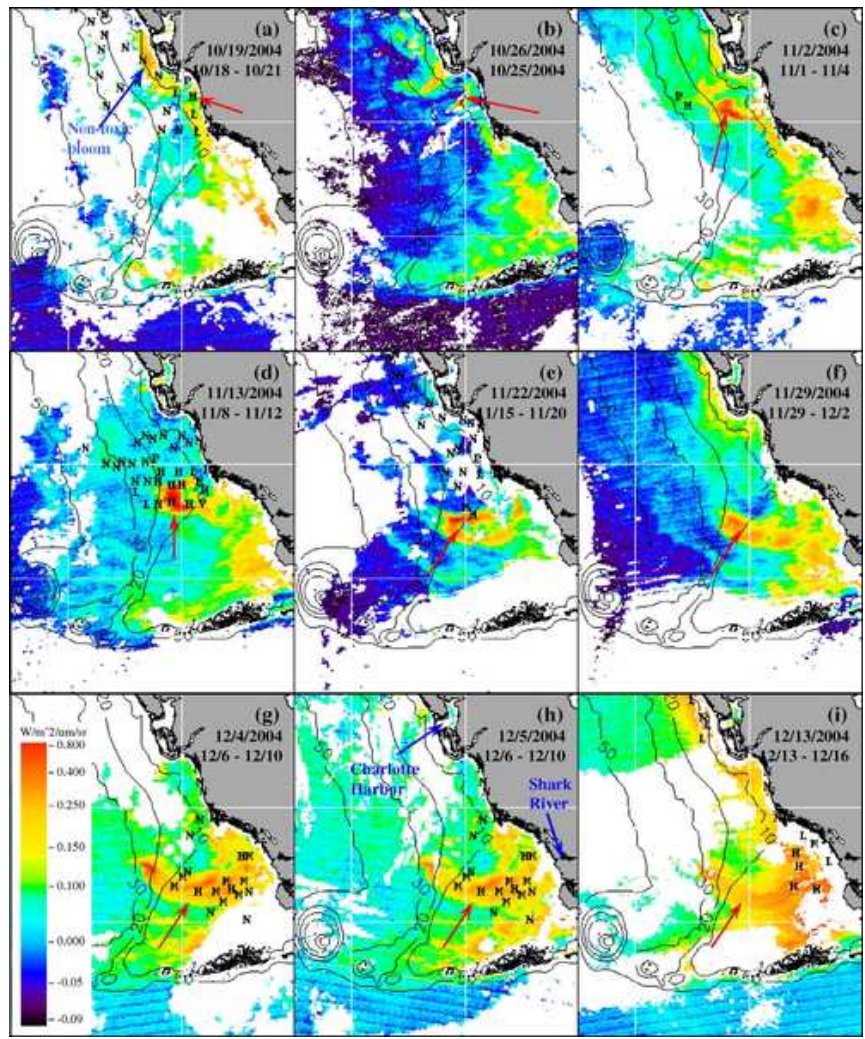}
\caption{Sequence of MODIS (Moderate Resolution Imaging
Spectroradiometer) fluorescence line height images showing the
progression of a harmful algal bloom produced by \emph{K.
brevis} (indicated with red arrows) on the southern portion of
the West Florida Shelf. Indicated parallels and meridians are
25$^{\circ}$N and 26$^{\circ}$N, and 82$^{\circ}$W and
83$^{\circ}$W, respectively. Superimposed on the images are
water sample analysis results from the Florida Fish and
Wildlife Research Institute. The second date on each image
indicates the in situ sample collection time. Letters represent
different \Kb concentrations in cell l$^{-1}$ as follows: N,
not present or below detection limit; P, present (smaller than
$10^3$); L, low (between $10^3$ and $10^4$); M, medium (between
$10^4$ and $10^5$); H, high (between $10^5$ and $10^6$); and V,
very high (larger than 106). Reprinted from \citet{Hu-etal-05}
with permission. Copyright Elsevier 2005.}
\label{bananaFig0}
\end{figure}

The goal of this paper is to demonstrate the utility of
simulated LCSs to help trace the early stages of the
development of HABs, and consequently to help acquiring a
better understanding of the environmental factors that may lead
to their development. We focus on a HAB that was observed on
the southern portion of the WFS during October--December 2004
(Figure \ref{bananaFig0}) \citep{Hu-etal-05} . Focus on this
HAB is placed both because of good event coverage and because
it preceded a anomalously active HAB season which extended over
most of 2005. The HAB was detected offshore using MODIS
(Moderate Resolution Imaging Spectroradiometer) fluorescence
line height imagery. The remote sensing detected high
concentrations of chlorophyll, and water sampling and
microscopic counts confirmed that this HAB contained mostly
\emph{K. brevis}. The early location of this HAB is traced here
using simulated LCSs, which carry critical information that
highly constrain fluid particle motion. The environmental
conditions that may have led to the development of the HAB in
question are then inferred using a nutrient--phytoplankton
population dynamics model. Our LCSs analysis and population
dynamics modeling effort are based on a simulation of the WFS
produced by the HYbrid-Coordinate Ocean Model (HYCOM).

\begin{figure}[t]
\end{figure}

\section{Lagrangian Coherent Structures}

A practical way to identify LCSs consists in computing direct
finite-time Lyapunov exponents (DFTLEs) \citep{Haller-01a,
Haller-02,Shadden-etal-05,Lekien-etal-05,Shadden-etal-06,
Mathur-etal-07,Lekien-etal-07,Green-etal-07}. The DFTLE is a
scalar that measures the finite-time average of the maximum
separation rate of pairs of passively advected fluid particles.
More specifically, the DFTLE is defined as
\begin{eqnarray}\label{DFTLE}
    \sigma_{t_0}^{\tau}(\bfx_0) &\coloneqq&
    \frac{1}{|\tau|}\ln\|\partial_{\bfx_0}\bfx(t_0 +
    \tau;\bfx_0,t_0)\|,
\end{eqnarray}
where $\|\,\| $ denotes spectral norm and $\bfx(t_0 +
\tau;\bfx_0,t_0)$ is the position at time $t_0 + \tau$ of a
fluid particle that at time $t_0$ was located at $\bfx_0$. The
latter is obtained by integrating the particle trajectory
equation
\begin{eqnarray}\label{dxdt}
    \dot{\bfx} &=& \bfu(\bfx,t),
\end{eqnarray}
where the overdot stands for time differentiation and
$\bfu(\bfx,t)$ is the fluid velocity.

Regions of maximum material line stretching produce local
maximizing curves or ``ridges'' in the DFTLE field. When the
DFTLE is computed by integrating particle trajectories backward
(forward) in time, $\tau < 0$ ($\tau > 0$), a ridge in the
DFTLE field corresponds to an attracting (repelling) LCS. The
repelling and attracting LCSs provide, respectively, a
generalization of the concepts of stable and unstable manifolds
of a hyperbolic (stagnation) point to the case of aperiodically
time-dependent velocity fields. The attracting and repelling
LCSs delineate, like a separatrix, the boundary between regions
with different flow characteristics which do not mix, and
constrain, respectively, the past and future history of
passively advected fluid particles.

\begin{figure}[t]
\centering
\includegraphics[width=39pc]{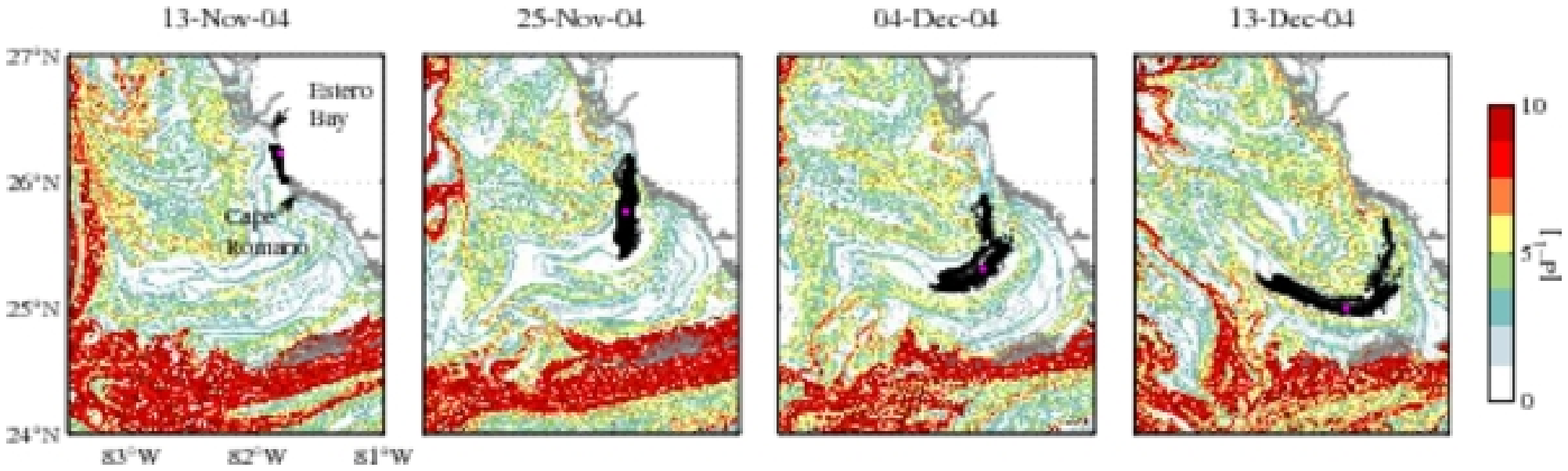}
\caption{Snapshots of direct finite-time Lyapunov exponent
(DFTLE) field (\ref{DFTLE}) with initial conditions on the
southern portion of the West Florida Shelf and computed
backward in time using surface currents produced by a HYCOM
(HYbrid-Coordinate Ocean model) simulation. Maximizing ridges
of backward-time DFTLE field indicate attracting Lagrangian
coherent structures (LCSs) or regions of maximum material line
stretching toward which fluid particles converge as time
progresses. These LCSs are hidden in the velocity field and
thus cannot be inferred visually from snapshots. The clouds of
black dots with the magenta dot immersed indicate simulated
passively advected particles. These were used, as explained in
the text, in tracing the early development of a simulated
harmful algal bloom (HAB), whose evolution is shown in Figure
\ref{bananaFig2}, which reproduces the main characteristics of
the observed HAB shown in Figure \ref{bananaFig0}.}
\label{bananaFig1a}
\end{figure}

\begin{figure}[t]
\centering
\includegraphics[width=39pc]{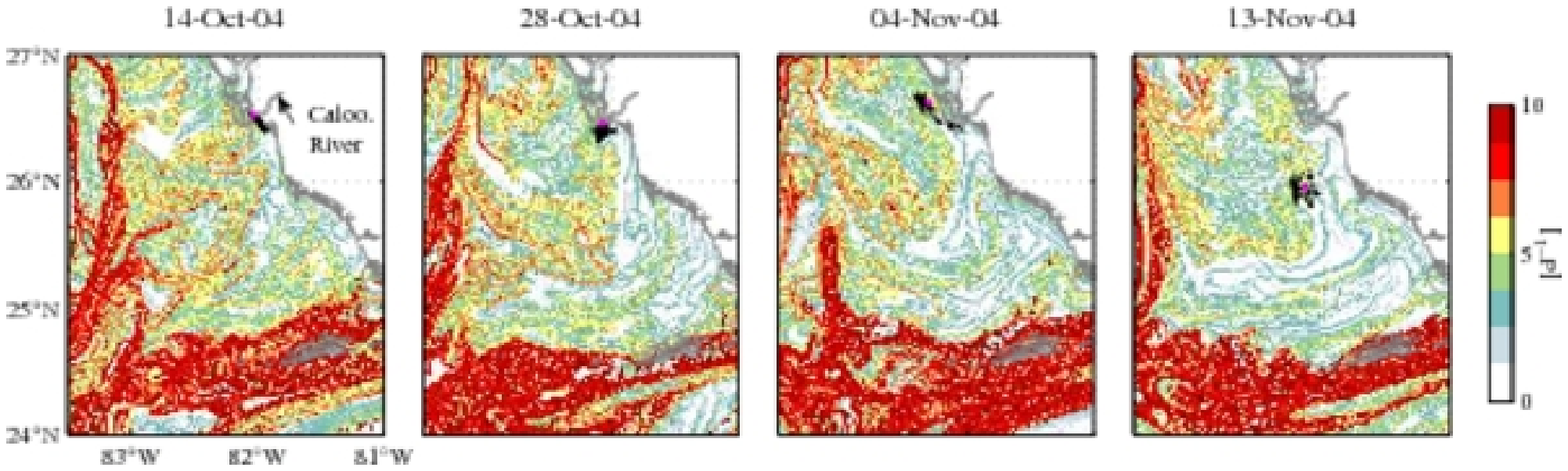}
\caption{As in Figure \ref{bananaFig1a}, but for a different set
of dates. Indicated in the left panel is the Caloosahatchee
River, which has been abbreviated as Caloo. River.}
\label{bananaFig1b}
\end{figure}

Numerical model output provides a flow description
$\mathbf{u}(\mathbf{x},t)$ that is suitable for use in the
identification of LCSs. In \citet{Olascoaga-etal-06} we
considered daily surface velocity fields extracted in the WFS
domain from a 1/25$^{\circ}$-resolution, free-running HYCOM
simulation of the Gulf of Mexico (GoM), itself nested within a
1/12$^{\circ}$-resolution Atlantic basin data assimilative
nowcast, which was generated at the U.S. Naval Research
Laboratory as part of the Global Ocean Data Assimilation
Experiment \citep{Chassignet-etal-07}. Using these surface
currents, in \citet{Olascoaga-etal-06} we identified LCSs on
the WFS that revealed the presence of a cross-shelf transport
barrier in approximately the same location as the western
boundary of the ``forbidden zone,'' which is a region on the
southernmost part of the WFS that was found not to be visited
by drifters that were released outside of the region
\citep{Yang-etal-99}. A highly convoluted portion of the
aforementioned transport barrier can be seen in Figures
\ref{bananaFig1a} and \ref{bananaFig1b}, which show on selected
days in 2004 DFTLE field snapshots computed by integrating
(\ref{dxdt}) backward in time, with $\tau = -60$ d. The portion
of the cross-shelf transport barrier is identified in this
figure as the DFTLE field ridges shown with intense red tones.
To construct this and the subsequent figure we considered
surface currents produced by a HYCOM simulation of the WFS
which, unlike that considered in \citet{Olascoaga-etal-06},
assimilated during the period 2004--2005 satellite-derived sea
surface height and temperature data, as well as available
vertical profile data in the GoM using the U.S. Navy Coupled
Ocean Data Assimilation system
\citep{Cummings-05,Hogan-etal-07}. The validity of the surface
circulation produced by this HYCOM simulation is supported by a
very good agreement of simulated LCSs with available drifter
trajectories on the dates of the simulation, which can be
accessed at the NOAA's Atlantic Oceanographic and
Meteorological Laboratory website
http://www.aoml.noaa.gov/sfros/drifters, and, as shown below,
the ability of the HYCOM simulation to produce surface currents
which lead to a simulated HAB with similar characteristics as
those of the event observed on the WFS in late 2004.

The focus in this paper is not on the above persistent, large
scale LCS which constitutes a barrier that inhibits transport
across the WFS. Rather, it is on other less persistent, smaller
scale LCSs, which lie on the shoreward side of the more
prominent cross-shelf transport barrier. The LCSs in question
are identified in Figures \ref{bananaFig1a} and
\ref{bananaFig1b} with the DFTLE field ridges with tones
ranging from less intense red through most intense blue in the
colorscale. Note, in particular, the presence of a
banana-shaped region bounded by LCSs of the latter type. The
banana-shaped region, which is most clearly evident on mid
December 2004 but can be traced in time from late October 2004
(Figure \ref{bananaFig1a}), coincides remarkably well with the
area spanned by the HAB described in \citet{Hu-etal-05} on 13
December 2004 (Figure \ref{bananaFig0}i). The noted agreement
between modeled LCSs and the boundary of the observed HAB
distribution is exploited in the following sections to trace
the early development of the observed HAB.

\section{Tracing the Early Development of the Harmful Algal Bloom}

In this section we trace the early development of the HAB
described in \citet{Hu-etal-05} using the simulated LCSs,
which, as explained above, carry critical information that
strongly constrain fluid particle motion. We thus concentrate
on the DFTLE fields shown in Figures \ref{bananaFig1a} and
\ref{bananaFig1b}. Depicted on these figures also are
instantaneous particle positions computed by integrating
(\ref{dxdt}) using the same HYCOM-based flow description
$\mathbf{u}(\mathbf{x},t)$ employed in computing the DFTLE
fields.

Consider first the particle positions lying within the
aforementioned banana-shaped region that resembles the area
occupied by the HAB in question on 13 December 2004 (Figure
\ref{bananaFig1a}, right panel). Immersed in the cloud of
particles, which are indicated with black dots, is one particle
indicated with a bold magenta dot that was used for reference
in tracing the early location of the observed HAB. We chose the
position of this fiducial particle to lie approximately
centered within the banana-shaped region on 13 December 2004.
Integrating backward in time (\ref{dxdt}) during 30 d with this
initial condition, we found that the the early location of the
HAB distribution in question could have restricted to a
nearshore area between Estero Bay and Cape Romano (Figure
\ref{bananaFig1a}, left panel). The integration time was chosen
so as to correspond to the time it typically takes for \Kb to
develop a HAB. Note that the fiducial particle remains for all
dates shown in Figure \ref{bananaFig1a} within the LCSs that
delimit the banana-shaped region. The particle positions
indicated with black dots were computed by integrating
(\ref{dxdt}) forward in time during 30 d with initial positions
surrounding that of the fiducial particle on 13 November 2004
in a nearshore region between Estero Bay and Cape Romano, which
lies within the LCSs that bound the banana-shaped region on
that date. Note that as time progresses from 13 November 2004
the particles fill the banana-shaped region while it stretches
and bends. Note also that during the relevant time interval the
particles approach, but do not traverse, the DFTLE ridges that
bound the banana-shaped region, which provides strong support
for our assertion that these DFTLE ridges are attracting LCSs.

Consider now the fiducial particle shown in the right panel of
Figure \ref{bananaFig1b}. This fiducial particle was located in
the center of the observed HAB distribution on 13 November
2004, which is situated north of the banana-shaped region in
the DFTLE field on that date. Integrating (\ref{dxdt}) backward
in time during 30 d with this initial position, we found that
the early location of the HAB distribution in question could
have restricted to an area near the mouth of the Caloosahatchee
River (Figure \ref{bananaFig1b}, left panel). Integrating
(\ref{dxdt}) forward in time during 30 d for a cloud of
particles with initial positions surrounding that of the
fiducial particle on 14 October 2004, we found that these
particles approximately covered the area spanned by the
observed HAB on 13 November 2004. We emphasize that the motion
of the particles is completely determined by the underlying
LCSs. Note, in particular, the LCS on 14 October 2004 which
appears to originate in the Caloosahatchee River mouth area,
right before the aforementioned banana-shaped region starts to
develop. Clearly, particles initially located near the
shoreline north of the Caloosahatchee River mouth will never
enter the banana-shaped region. Contrarily, these particles
will disperse over a region characterized by a highly intricate
underlaying LCS.

In the following section we consider a simplified population
dynamics model which reproduces the observed HAB's main
characteristics. The population dynamics model considers
initial nutrient sources located in the two regions of
potential early HAB development estimated above based on the
information contained in the modeled LCSs.

\section{Population Dynamics Model}

A number of biological and physical factors contribute to the
initiation, growth, maintenance, and demise of HABs produced by
\emph{K. brevis}. We consider nutrient supply and surface ocean
circulation constraints as the most important factors.

\begin{figure}[t]
\centering
\includegraphics[width=39pc]{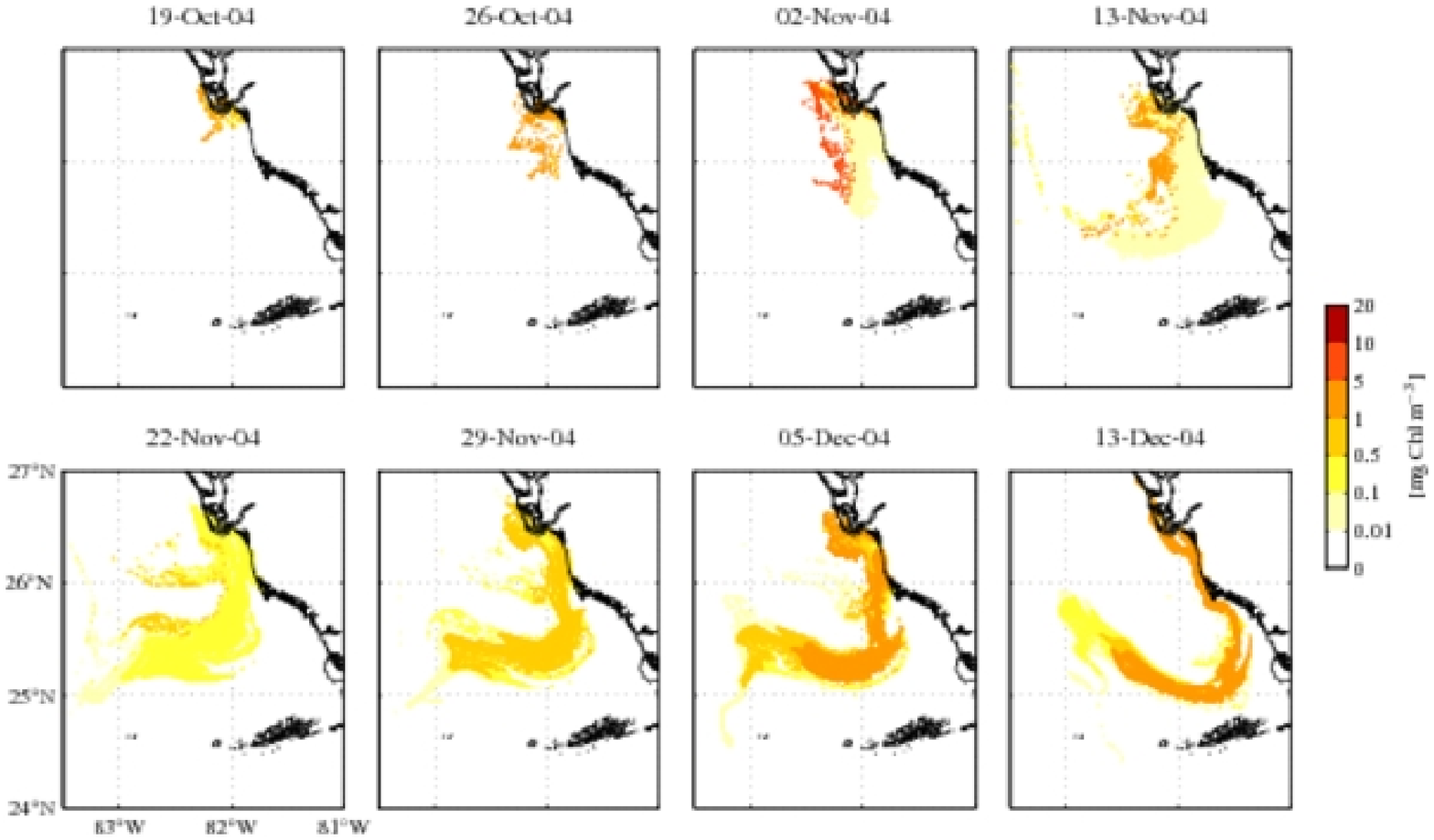}
\caption{Series of snapshots of the evolution of a harmful algal
bloom (HAB) produced by \emph{Karenia brevis} on the southern
West Florida Shelf as simulated according to (\ref{NP}) using a
HYCOM (HYbrid-Coordinate Ocean model) based surface current
description and the fluid particle motion information carried
in the Lagrangian coherent structures shown in Figures
\ref{bananaFig1a} and \ref{bananaFig1b}. The simulated HAB
reproduces the main features of the observed HAB shown in Figure
\ref{bananaFig0}.} \label{bananaFig2}
\end{figure}

Consistent with these assumptions, we formulate a population
dynamics model in which the nitrogen concentration is
determined in two compartments, nutrients ($N$) and
phytoplankton ($P$), which is partly justified by the lack of
\Kb grazers \citep{Steidinger-73}. We further assume that these
densities are passively advected by surface currents, which are
described by the HYCOM simulation of the WFS. The equations
describing the evolution of $N$ and $P$ take the form
\citep{Riley-63,Wroblewski-OBrien-76,
Wroblewski-etal-88}\begin{mathletters}\label{NP}
\begin{eqnarray}
    \partial_t N + \bfu \cdot \nabla N & = &
      - \mu \frac{N}{N + N_0} P + \varepsilon \sigma P, \\
    \partial_t P + \bfu \cdot \nabla P & = &
      \mu \frac{N}{N + N_0} P - \sigma P.
\end{eqnarray}
\end{mathletters}The nutrient-limited phytoplankton growth
is assumed to have a maximum rate $\mu = 0.4$ d$^{-1}$ and a
half-saturation constant $N_0 = 0.5$ mmol N m$^{-3}$. The
linear phytoplankton loss is parameterized by $\sigma = 0.1$
d$^{-1}$. The assumed proportion of recyclable nutrients in
dead phytoplankton $\varepsilon = 0.1$. These parameter values,
which are within the range of those considered in earlier works
\citep{Steidinger-etal-98,Walsh-Steidinger-01,Walsh-etal-01,
Walsh-etal-02,Walsh-etal-03}, result in a net population growth
rate of the order of 0.25 d$^{-1}$ in agreement with field
observations \citep{vanDolah-Leighfield-99}.

Equations (\ref{NP}) are solved using the method of the
characteristics by tracing $10^4$ particles, which are released
every other day on a uniformly distributed lattice extending
from 27$^{\circ}$N to 26$^{\circ}$N and from the shoreline to
the 10-m isobath. Attached to each particle initially is a
$P$-value of 0.01 mg Chl m$^{-3}$ (the conversion equivalence
for \Kb is assumed be 1 mg Chl m$^{-3} = 0.38$ mmol N m$^{-3}$
\citep{Shanley-Vargo-93}), which is a typical background \Kb
concentration value on the WFS \citep{Geesey-Tester-93}. Except
for those particles released near the shoreline, attached to
each particle initially is an $N$-value of 0.1 mmol N m$^{-3}$,
which corresponds to a normal oligotrophic condition on the WFS
\citep{Masserini-Fanning-00}. Attached to each particle
released on 1 October 2004 (start date of the population
dynamics model simulation) near the shoreline in the vicinity
of the Caloosahatchee River initially is an $N$-value of 15
mmol N m$^{-3}$, which decays following a Gaussian function to
oligotrophic initial $N$-values toward 13 December 2004 (end
date of the population dynamics model simulation) for particles
released on those intermediate dates. Attached to each particle
released on 13 November 2004 near the shoreline between Estero
Bay and Cape Romano initially is an $N$-value of 5 mmol N
m$^{-3}$, which decays to oligotrophic initial $N$-values
toward the start and end dates of the population dynamics model
simulation for particles released on those intermediate dates.

Figure \ref{bananaFig2} shows a sequence of snapshots of the
evolution of a simulated phytoplankton distribution which
evolves according to (\ref{NP}). Note the good qualitative
agreement with the observed HAB distributions shown in Figure
\ref{bananaFig0}, which confirms the utility of the information
carried in LCSs to help identify regions of potential HAB
initiation.

The results of the population dynamics model simulation are not
sensitive to small variations of model parameter values or the
addition of a small amount of diffusion in (\ref{NP}). However,
they are sensitive to the location of the nutrient sources,
which were revealed in the LCSs analysis, and the
characteristics of these sources, which may be understood as
follows. For particles released in the vicinity of the
Caloosahatchee River, the assumed nonoligotrophic initial
nutrient concentrations may be associated with nutrient-rich
water from land runoff, which are believed to play an important
role in HAB development \citep{Brand-Compton-07}. The assumed
nutrient concentration values and time dependence roughly
adhere to measurements of, respectively, nutrient content and
flow discharge near the mouth of the Caloosahatchee River for
the dates of the population dynamics model simulation. Nutrient
content data and flow discharge time series are available at
the U.S. Geological Survey website
http://waterdata.usgs.gov/nwis/nwisman/\allowbreak
?site\_no=02292900\allowbreak\&agency\_cd=USGS. The large
Caloosahatchee River flow discharges in 2004 can be attributed
to a very active hurricane season. For particles released
between Estero Bay and Cape Romano, the assumed nonoligotrophic
initial nutrient concentrations may be associated with an
upwelled intrusion of nutrient-rich GoM slope water on the WFS,
which is consistent with the HYCOM simulation considered here
and AVHRR (Advanced Very High Resolution Radiometer) sea
surface temperature maps, e.g.,
http://marine.rutgers.edu/mrs/show/\allowbreak
?file=../regions/floridacoast/sst/noaa/2004/img/041111.316.0318.n17.jpg.
In this case, the assumed time dependence simulates the
intensity variation of the upwelling event for the dates of the
population dynamics model simulation.

\section{Summary and Conclusions}

In this paper we have traced the early stages of the
development of a harmful algal bloom (HAB) produced by
\emph{Karenia brevis} which was observed on the southern
portion of the West Florida Shelf (WFS) during
October--December 2004. This HAB was detected offshore using
ocean color imagery and field sampling. The early location of
this HAB was traced with the aid of simulated Lagrangian
coherent structures (LCSs). The factors that may have led to
the development of this HAB were then inferred using a
population dynamics model. The LCSs determine fluid particle
pathways hidden in the velocity field which highly constrained
the evolution of the HAB in question. To perform the LCS
computation and population dynamics modeling we made use of a
surface velocity field description provided by a HYCOM
(HYbrid-Coordinate Ocean Model) simulation of the WFS. With the
aid of the simulated LCSs, which were identified as ridges in
direct finite-time Lyapunov exponent fields, two nearshore
nutrient sources were identified as responsible for the
development of a simulated HAB which reproduced the main
features of the event observed on the WFS in 2004. One source,
which can be associated with an upwelled intrusion of
nutrient-rich water on the WFS, was located in an area near the
coastline between Estero Bay and Cape Romano. The other source,
which can be associated with nutrient-rich water from land
runoff, was located at the mouth of the Caloosahatchee River.
The latter finding is consistent with the hypothesis which
considers human-related nutrient loading in coastal waters as a
critical factor leading to HAB development. The results of our
work demonstrate that LCS simulation in combination with
population dynamics modeling can greatly enhance our ability to
understand the early stages of HAB development.

\begin{acknowledgments}
We are thankful to G. Halliwell and O. Smedstad for proving the
HYCOM model output, M. Brown and S. Smith for critically
reading the manuscript, and I. Rypina and I. Udovydchenkov for
the benefit of discussions on dynamical systems theory. MJO was
supported by the NSF grant CMG-0417425, the PARADIGM
NSF/ONR-NOPP grant N000014-02-1-0370, the NSF grant OCE0432368,
and the NIEHS grant P50 ES12736. FJBV and HK were supported by
the NSF grant CMG-0417425. LEB was supported by the NSF grant
OCE0432368 and the NIEHS grant P50 ES12736.
\end{acknowledgments}

\bibliographystyle{agu04}

\end{article}

\end{document}